\documentclass[10pt,letterpaper,twocolumn,showpacs,prl,aps]{revtex4-1} 

\usepackage{amsmath}  
\usepackage{amsfonts} 
\usepackage{graphicx} 
\usepackage{amssymb}
\usepackage{url}
\usepackage{hyperref}

\begin{document}

\title{Exact Single-Scale Outer Solution of the Abrikosov Vortex in the Extreme Type-II Limit}

\author{Eugene B. Kolomeisky}

\affiliation
{Department of Physics, University of Virginia, P. O. Box 400714, Charlottesville, Virginia 22904-4714, USA}

\date{\today}

\begin{abstract}
We determine the exact outer structure of the Abrikosov vortex in the extreme type-II limit, which occurs when the Ginzburg-Landau parameter $\kappa$ diverges. In this limit, Ginzburg-Landau theory simplifies, outside a shrinking core, to a closed nonlinear theory for the superfluid velocity subject to an algebraic density constraint. The resulting solution is asymptotically exact everywhere outside the vanishing vortex core, demonstrating that both magnetic field and superconducting density vary on the length scale of the London penetration depth. This establishes that the conventional two-length-scale picture of the vortex does not hold in the $\kappa\gg 1$ limit.    
  
\end{abstract}


\maketitle

The Abrikosov vortex is one of the central objects of superconductivity \cite{Abrikosov1,Abrikosov2,de Gennes,LL9} and gauge field theory \cite{Abelian,Rubakov}.  Within Ginzburg-Landau (GL) theory its structure is conventionally described in terms of two intrinsic length scales: the coherence length $\xi$ and the London penetration depth $\lambda$.   This two-scale structure underlies the distinction between type-I and type-II superconductors and is deeply embedded in textbook treatments.  Depending on the value of the GL parameter $\kappa=\lambda/\xi$, superconductors are either type-I or type-II materials.

In this work we reexamine the vortex in the extreme type-II limit $\kappa\rightarrow\infty$. We show that in this singular limit the GL equations reduce, outside a core whose size shrinks as $\kappa^{-1}$, to a closed nonlinear theory for the gauge-invariant superfluid velocity. The superconducting density becomes algebraically slaved to the velocity field, and the vortex exterior is determined exactly by a single length scale, the London penetration depth. The resulting structure is not merely a large-$\kappa$ asymptotic approximation but the exact outer solution of the limiting theory.  

Previous large-$\kappa$ treatments \cite{Abrikosov1,Abrikosov2,de Gennes,LL9} have derived asymptotic vortex forms through matching procedures. In contrast, the reduction presented here reveals that the outer problem becomes exactly solvable once expressed in terms of the velocity field alone. Variations of the superconducting density are not frozen, as in the London model, but retained through an exact algebraic constraint that the limiting nonlinear theory inherits from GL theory.  

Thus, while finite-$\kappa$ GL theory formally contains two intrinsic length scales, the extreme type-II limit exhibits a qualitatively different structure: outside a vanishing core, the vortex reorganizes into a single-scale configuration governed by an exact solution.

The two-length-scale structure of GL theory forms the basis of the London description of the mixed state in extreme type-II superconductors developed by Abrikosov \cite{Abrikosov1,Abrikosov2} and further elaborated in Refs.~\cite{de Gennes,Matricon,Fetter,Sarma}. The underlying assumption is that variations of the superconducting density are confined to narrow vortex cores of radius of order of the coherence length. Since over a wide range of external magnetic fields these cores occupy only a small fraction of the sample volume when $\kappa\gg 1$, the superconducting density may be approximated by its bulk value almost everywhere. The GL equations then reduce to the linear London equation, whose vortex solutions are necessarily singular at the vortex positions.

The resulting physical picture became widely accepted despite the fact that the original work \cite{Abrikosov1,Abrikosov2} did not explicitly analyze the full single-vortex solution of the GL equations. 
\begin{figure}
\begin{center}
\includegraphics[width=1\columnwidth]{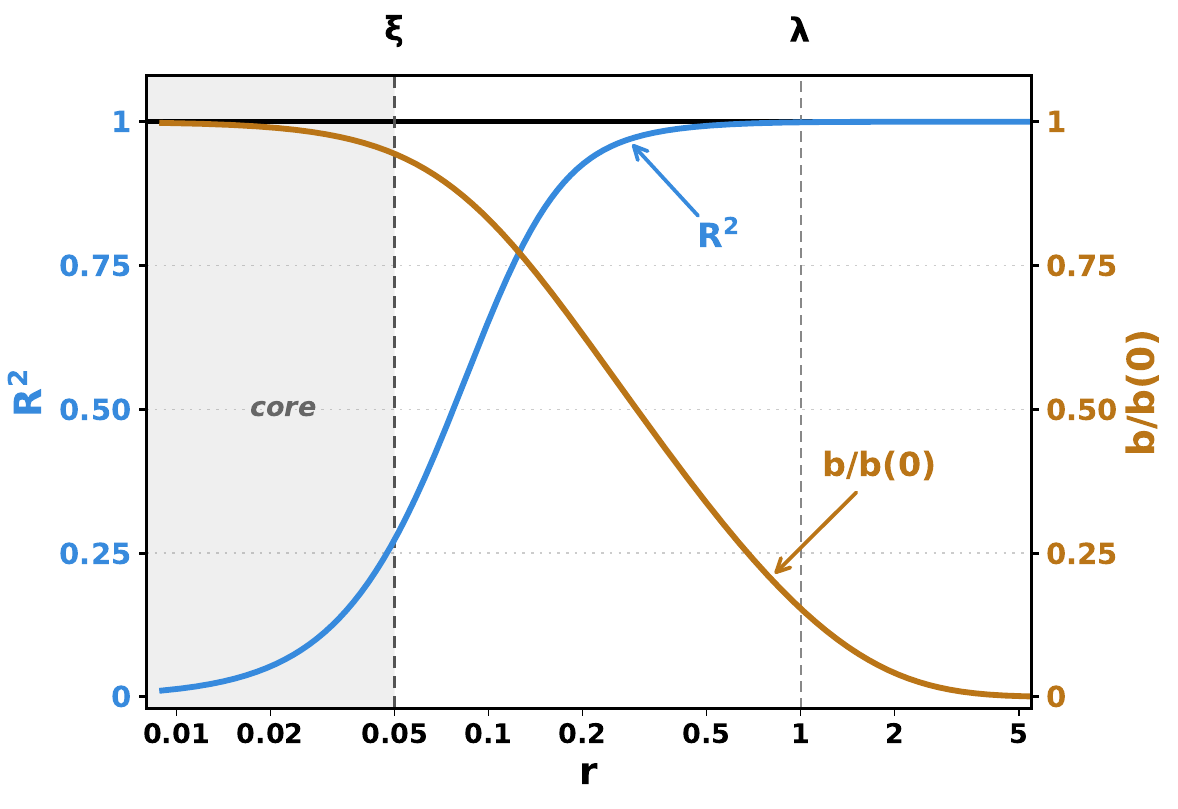}
\caption{Rescaled superconducting density $R^{2}$ and rescaled magnetic induction $b/b(0)$ as functions of the distance $r$ (in units of the London penetration depth $\lambda$, logarithmic scale) from the vortex axis according to the GL equations (\ref{GL1_vortex}) and (\ref{GL2_vortex}) and $\kappa=20$.  The shaded region $r\lesssim\xi$, where $\xi$ is the coherence length, is the vortex core.  Textbook illustrations \cite{de Gennes, LL9} suggest distinct recovery scales for magnetic induction and superconducting density. In the extreme type-II limit, however, the exact outer solution, Eqs.(\ref{magnetic_induction_exact}) and (\ref{superconducting_density_exact}), shows that both recover on the same scale outside a shrinking core.}
\label{GL_vortex_kappa20}
\end{center}
\end{figure}
Figure~\ref{GL_vortex_kappa20} shows the numerical solution of the GL  equations (\ref{GL1_vortex}) and (\ref{GL2_vortex}) for a single-flux-quantum vortex at $\kappa=20$. Outside the narrow vortex core, both the superconducting density and the magnetic induction recover toward their bulk values over distances of order of the penetration depth rather than the coherence length. The figure therefore already suggests that, in the extreme type-II regime, the outer vortex structure is governed by a single macroscopic scale. Similar numerical behavior has been documented previously in Refs.~\cite{Fetter,Sarma}.     

The breakdown of the conventional two-length-scale picture in the $\kappa\rightarrow\infty$ limit is analogous to the Born-Oppenheimer approximation in quantum mechanics \cite{BO}. In systems with strongly separated scales, fast degrees of freedom do not evolve independently but adiabatically follow the slow ones. Likewise, although the superconducting density by itself would recover on the short scale $\xi$, in the extreme type-II limit it becomes slaved to the superfluid velocity through Eq.~(\ref{slaving_condition}) and therefore inherits the longer scale $\lambda$ governing the velocity field.   

Employing the GL units \cite{Abrikosov1,Abrikosov2,LL9,Fetter} and the amplitude-phase representation of the superconducting order parameter $\Psi(\mathbf{r})=R(\mathbf{r})e^{i\theta(\mathbf{r})}$ where $R^{2}$ is the superconducting density and $\theta$ is the phase of the order parameter, the GL free energy functional below the temperature of superconducting transition can be written in the form
\begin{equation}
\label{GL_free_energy_dimensionless}
f[\mathbf{a},R,\theta]=\int \Bigg\{ \frac{(\nabla R)^{2}}{\kappa^{2}} +R^{2} \mathbf{v}^{2}
+\frac{1}{2}(R^{2}-1)^{2}+(\nabla\times \mathbf{a})^{2}\Bigg\}dV,   
\end{equation}
\begin{equation}
\label{superconducting_velocity_GL_units}
\mathbf{v}=\frac{1}{\kappa}\nabla\theta-\mathbf{a}
\end{equation}
where $\mathbf{v}$ is the superfluid velocity and $\mathbf{a}$ is the vector potential determining the magnetic induction $\mathbf{b}=\nabla\times\mathbf{a}$.  In the GL units the length is measured in units of the London penetration depth so that the coherence length is $\xi=1/\kappa$.  The free energy density in Eq.(\ref{GL_free_energy_dimensionless}) vanishes for $R^{2}=1$, $\mathbf{v}=0$ and $\mathbf{b}=0$, the lowest free energy state corresponding to bulk uniform superconductor without supercurrent. 

Taking the curl of the expression for the superfluid velocity (\ref{superconducting_velocity_GL_units}) determines the magnetic induction \cite{Abrikosov1,Abrikosov2,LL9}
\begin{equation}
\label{magnetic_field}
\textbf{b}=-\nabla\times\mathbf{v}+\frac{2\pi}{\kappa}\delta^{(2)}(\mathbf{r})\mathbf{e}_{z}
\end{equation}
where we specified to the case of a single straight Abrikosov vortex carrying one flux quantum.  The vector $\mathbf{r}$ represents the distance from the vortex and $\mathbf{e}_{z}$ is the unit vector along the vortex axis chosen to be the $z$-axis of the cylindrical coordinate system.  Cylindrical symmetry of the vortex problem implies that the magnetic field is pointing along the $z$-axis, $\mathbf{b}=b\mathbf{e}_{z}$, and that the velocity field is azimuthal, $\mathbf{v}=v\mathbf{e}_{\varphi}$ where $\mathbf{e}_{\varphi}$ is the azimuthal unit vector.  Since the magnetic induction is finite at the vortex axis, in its vicinity the velocity field satisfies the singular boundary condition \cite{Abrikosov1,Abrikosov2,LL9}
\begin{equation}
\label{singular_velocity}
v(\mathbf{r}\rightarrow 0)=\frac{1}{\kappa r}.
\end{equation} 

Following Abrikosov \cite{Abrikosov1,Abrikosov2}, we formulate the theory directly in terms of the gauge-invariant superfluid velocity field (\ref{superconducting_velocity_GL_units}). From a modern perspective this corresponds to a unitary-gauge transformation in which the phase degree of freedom is absorbed into the vector potential, analogous to the construction later used in discussions of the Anderson-Higgs mechanism \cite{A,H1,H2,H3,H4,H5}. In the presence of a vortex the gauge transformation (\ref{superconducting_velocity_GL_units}) is singular on the vortex axis, reflecting quantization of magnetic flux. In this representation the topological content of the theory is retained through the boundary condition (\ref{singular_velocity}), rather than through a delta-function singularity in (\ref{magnetic_field}). The magnetic induction therefore satisfies $\mathbf{b}=-\nabla\times\mathbf{v}$, and the GL free energy (\ref{GL_free_energy_dimensionless}) reduces to a functional of the order-parameter amplitude $R$ and the superfluid velocity field $\mathbf{v}$ only:
\begin{equation}
\label{GL_free_energy_velocity_gauge}
f[\mathbf{v},R]=\int \left [\frac{(\nabla R)^{2}}{\kappa^{2}} +R^{2} \mathbf{v}^{2}
+\frac{1}{2}(R^{2}-1)^{2}+\mathbf{b}^{2}\right ]dV.
\end{equation} 
Minimizing with respect to $R$ and $\mathbf{v}$ we obtain the GL field equations
\begin{equation}
\label{GL1}
\frac{1}{\kappa^{2}}\nabla^{2}R+(1-R^{2}-\mathbf{v}^{2})R=0,
\end{equation}
\begin{equation}
\label{GL2}
\nabla\times\mathbf{b}=-\nabla\times(\nabla\times\mathbf{v})=R^{2}\mathbf{v}.
\end{equation}
First of these is a nonlinear Schr\"odinger equation while the second is Amp\`ere's law.  

Taking the $\kappa\rightarrow\infty$ limit in Eq. (\ref{GL_free_energy_velocity_gauge}) simplifies it to
\begin{equation}
\label{GL_free_energy_infinite_kappa}
f[\mathbf{v},R]=\int \left [R^{2} \mathbf{v}^{2}+\frac{1}{2}(R^{2}-1)^{2}+(\nabla\times \mathbf{v})^{2}\right ]dV.
\end{equation}
Further, minimizing with respect to the superconducting density $R^{2}$, or, equivalently, taking the $\kappa\rightarrow\infty$ limit in Eq.(\ref{GL1}), we obtain a relationship
\begin{equation}
\label{slaving_condition}
R^{2}=1-\mathbf{v}^{2}
\end{equation}
hereafter called the slaving condition.  Since $R^{2}\ge 0$, the superfluid  velocity is constrained by the inequality $|\mathbf{v}|\le 1$;  $v=1$ is the critical velocity to destroy superconductivity, the exact property of the $\kappa\rightarrow\infty$ limit of GL theory.  The transition from the differential GL equation (\ref{GL1}) to the algebraic slaving condition (\ref{slaving_condition}) illustrates singular nature of the $\kappa\rightarrow\infty$ limit.  

Substituting the slaving condition (\ref{slaving_condition}) into the free energy functional (\ref{GL_free_energy_infinite_kappa}) and Amp\`ere's law (\ref{GL2}) transforms them into 
\begin{equation}
\label{AL_free_energy}
f[\mathbf{v}]=\int \left [(\nabla\times \mathbf{v})^{2}+\mathbf{v}^{2}-\frac{1}{2}\mathbf{v}^{4}\right ]dV,~~|\mathbf{v}|\le 1,
\end{equation}
\begin{equation}
\label{nonlinear_Ampere}
\nabla\times\mathbf{b}=-\nabla\times(\nabla\times\mathbf{v})=(1-\mathbf{v}^{2})\mathbf{v}.
\end{equation}

To summarize, in the $\kappa\rightarrow\infty$ limit GL theory simplifies to a closed nonlinear theory for the superfluid velocity field $\mathbf{v}$ only.  The magnetic induction is determined by the velocity, $\mathbf{b}=-\nabla\times\mathbf{v}$, and the superconducting density $R^{2}$ is slaved to the velocity via the relationship (\ref{slaving_condition}).  
Central equations of the theory (\ref{AL_free_energy}) and (\ref{nonlinear_Ampere}) along with the slaving condition (\ref{slaving_condition}) do not contain the GL parameter $\kappa$ making it clear that the $\kappa\rightarrow\infty$ limit of GL theory is single-scale:  both magnetic induction and superconducting density vary on the scale of the London penetration depth.  The GL parameter $\kappa$ only enters via the boundary condition (\ref{singular_velocity}).   
 
The $\nabla\times\mathbf{v}=-\mathbf{b}=0$ part of the free energy density in (\ref{AL_free_energy}) has a minimum at $\mathbf{v}^{2}=0$ corresponding, via the slaving condition (\ref{slaving_condition}), to  $R^{2}=1$;  this is the superconducting phase.  Likewise, the free energy density in (\ref{AL_free_energy}) has a maximum at $\mathbf{v}^{2}=1$ corresponding to $R^{2}=0$;  this is the normal phase.  In the $|\mathbf{v}|\ll1$ limit the free energy (\ref{AL_free_energy}) reduces to that of the London theory \cite{Abrikosov1,Abrikosov2,de Gennes, LL9}.

Accounting for cylindrical symmetry of the Abrikosov vortex, GL equations (\ref{GL1}) and (\ref{GL2}) acquire the form \cite{Abrikosov1,Abrikosov2}
\begin{equation}
\label{GL1_vortex}
\frac{1}{\kappa^{2}r}\frac{d}{dr}\left (r\frac{dR}{dr}\right )+(1-R^{2}-v^{2})R=0,
\end{equation}  
\begin{equation}
\label{GL2_vortex}
-\frac{db}{dr}=\frac{d}{dr}\left [\frac{1}{r}\frac{d}{dr}(rv)\right ]=R^{2}v.
\end{equation}
Vortex solution to these equations satisfies the boundary conditions:  $R(r\rightarrow\infty)=1$,  $v(r\rightarrow\infty)=0$, $R(r\rightarrow 0)=0$ as well as Eq.(\ref{singular_velocity}).   In the $\kappa\rightarrow\infty$ limit they reduce to a single nonlinear equation for the velocity field (\ref{nonlinear_Ampere}):
\begin{equation}
\label{semiclassical_v}
\frac{d}{dr}\left [\frac{1}{r}\frac{d}{dr}(rv)\right ]=(1-v^{2})v.
\end{equation}
In the $v^{2}\ll 1$ limit this equation becomes linear whose solution vanishing at $r$ large is $v\propto K_{1}(r)$ where hereafter $K_{\nu}(r)$ stands for the modified Bessel function of the second kind of order $\nu$ \cite{NIST}.  If we write 
\begin{equation}
\label{velocity_exact}
v=\frac{1}{\kappa}K_{1}(r),
\end{equation}
then the small $r$ limit, Eq.(\ref{singular_velocity}), will be also captured since $K_{1}(r\rightarrow 0)=1/r$ \cite{NIST}.  Then Eq.(\ref{velocity_exact}) correctly interpolates between small and large distance behavior of the velocity field.  We further observe that in the $\kappa\rightarrow\infty$ limit the solution (\ref{velocity_exact}) becomes exact.  Indeed, substituting Eq.(\ref{velocity_exact}) in the full nonlinear equation (\ref{semiclassical_v}) we see that the latter, in the $\kappa\rightarrow\infty$ limit, is satisfied identically.  

The magnetic induction and the superconducting density (\ref{slaving_condition}) can be inferred from the velocity field (\ref{velocity_exact}) as
\begin{equation}
\label{magnetic_induction_exact}
b=-\frac{1}{r}\frac{d}{dr}(rv)=\frac{1}{\kappa}K_{0}(r),
\end{equation}
\begin{equation}
\label{superconducting_density_exact}
R^{2}=1-\frac{K_{1}^{2}(r)}{\kappa^{2}}.
\end{equation}
The latter vanishes at $r=1/\kappa$, the coherence length, where the superfluid velocity (\ref{velocity_exact}) reaches its critical value $v=1$;  the $r<1/\kappa$ region is identified with the normal, $R=0$, vortex core.  We conclude that in the $\kappa\rightarrow\infty$ limit the exact outer, $r>1/\kappa$, structure of the Abrikosov vortex is given by Eqs.(\ref{velocity_exact}), (\ref{magnetic_induction_exact}), and (\ref{superconducting_density_exact}).  

The line energy of the vortex, following from Eq.(\ref{GL_free_energy_velocity_gauge}), is
\begin{equation}
\varepsilon=2\pi\int_0^\infty\left[\frac{1}{\kappa^{2}}\left (\frac{dR}{dr}\right )^{2}+R^2v^2+\frac12\left(1-R^2\right)^2+b^2\right]rdr. 
\label{eq:eps-GL}
\end{equation}
In the velocity theory the core, $r<1/\kappa$, is normal and carries the condensation energy
$\tfrac12$ per unit volume, that is, $\pi/2\kappa^2$ per unit length. Outside it, Eqs.(\ref{slaving_condition}), (\ref{velocity_exact}) and (\ref{magnetic_induction_exact}) hold, and the integrand of Eq.~(\ref{eq:eps-GL}) reduces to that of Eq.(\ref{AL_free_energy}), $b^2+v^2-\tfrac12v^4$. Hence the vortex line energy according to the nonlinear velocity theory is
\begin{eqnarray}
\varepsilon_{\rm v}^{(\infty)}&=&\frac{\pi}{2\kappa^2}+\frac{2\pi}{\kappa^2}\int_{1/\kappa}^{\infty}\left (K_0^2+K_1^2-\frac{K_1^4}{2\kappa^2}\right )rdr\nonumber\\
&=&\frac{2\pi}{\kappa^2}\left [\frac{1}{4}+\ln(2\kappa)-\gamma-\frac{1}{4}\right ]+O\!\left(\frac{\ln^2\kappa}{\kappa^4}\right).
\label{eq:eps-v}
\end{eqnarray}
The quadratic terms integrate exactly, $\frac{d}{dr}\left[rK_0K_1\right]=-r\left(K_0^2+K_1^2\right)$, and
give $\ln(2\kappa)-\gamma$ where $\gamma=0.57721...$ is Euler's constant \cite{NIST}; the quartic term contributes $-\tfrac14$, all of it from $r\sim1/\kappa$, and is cancelled by the condensation energy of the core supplying $+\tfrac14$.

Equation (\ref{eq:eps-v}) fixes the leading logarithm and the screening part of the constant, $\ln2-\gamma$, but not the contribution of the core:  the true amplitude $R$ is smooth rather than sharp, and the gradient energy contribution $(dR/dr)^{2}/\kappa^2$ into the line energy, absent from the $\kappa\to\infty$ theory, is of order $1/\kappa^{2}$.  To account for this we note that the magnetic flux $\sim\pi L^{2}|\ln L|/\kappa$ enclosed within a cylinder of radius $L\ll 1$ (see Eq. (\ref{magnetic_induction_exact})) is much smaller than the flux quantum $2\pi/\kappa$.  Therefore on such scales magnetic effects are negligible, $v=1/(\kappa r)$ and the Abrikosov vortex becomes the Ginzburg-Pitaevskii (GP) vortex of a neutral superfluid \cite{GP}.  Indeed, substituting $v=1/(\kappa r)$ into the first GL equation (\ref{GL1_vortex}) yields the GP equation for the amplitude
\begin{equation}
 \frac{1}{x}\frac{d}{dx}\left (x\frac{dR}{dx}\right )+\left (1-R^{2}-\frac{1}{x^{2}}\right )R=0
\label{GP_equation}
\end{equation}
where we employed the units of the coherence length, $r=x/\kappa$. The line energy of the GP vortex in a cylinder of radius $L\gg 1/\kappa$, that can be inferred from Eq.(\ref{eq:eps-GL}) by setting $v=1/(\kappa r)$ and $b=0$, is \cite{GP,Pethick}
\begin{eqnarray}
\varepsilon_{\rm GP}(L)&=&\frac{2\pi}{\kappa^{2}}\int_0^{\kappa L}\left[\left (\frac{dR}{dx}\right )^{2}+\frac{R^{2}}{x^{2}}+\frac12\left(1-R^2\right)^2\right]x dx\nonumber\\
&=&\frac{2\pi}{\kappa^{2}}\left [\ln(1.464\kappa L)+\frac{1}{4(\kappa L)^2}+\dots \right ]
\label{GP_energy}
\end{eqnarray}
where the $1/[4(\kappa L)^{2}]$ term is obtained by employing the far field limiting solution to the GP equation (\ref{GP_equation}), $R(x\gg 1)=1-1/(2x^{2})-9/(8x^{4})+\dots$. 

On the other hand, the velocity theory, with $R=0$ for $x<1$ and $R^{2}=1-1/x^{2}$ outside, Eq.(\ref{eq:eps-v}), gives
\begin{eqnarray}
\varepsilon_{\rm v}(L)&=&\frac{2\pi}{\kappa^{2}}\left [\frac{1}{4}+\int_{1}^{\kappa L}\left (\frac{1}{x}-\frac{1}{2x^{3}}\right )dx\right ]\nonumber\\
&=&\frac{2\pi}{\kappa^{2}}\left [\ln(\kappa L)+\frac{1}{4(\kappa L)^{2}}\right ].
\label{eq:E-v}
\end{eqnarray}
Both descriptions are valid in the overlap region $1/\kappa\ll r\ll1$,
and there they agree: for $x\gg1$ the GP solution approaches $R^2=1-1/x^{2}$, and the energy
densities of Eqs.~(\ref{GP_energy}) and (\ref{eq:E-v}) coincide, including the quartic term
that gives rise to $1/[4(\kappa L)^2]$; the gradient energy enters only at order $x^{-5}$.

Choosing the radius $L$ inside the overlap region, $1/\kappa\ll L\ll1$, we may then compute the
line energy within the cylinder of radius $L$ from the GP solution and the energy outside it from
the velocity theory:
\begin{equation}
\varepsilon=\varepsilon_{\rm GP}(L)+\left[\varepsilon_{\rm v}^{(\infty)}-\varepsilon_{\rm v}(L)\right]
=\varepsilon_{\rm v}^{(\infty)}+\left[\varepsilon_{\rm GP}(L)-\varepsilon_{\rm v}(L)\right].
\label{eq:matching}
\end{equation}
By Eqs.~(\ref{GP_energy}) and (\ref{eq:E-v}), the difference,
\begin{equation}
\varepsilon_{\rm GP}(L)-\varepsilon_{\rm v}(L)=\frac{2\pi}{\kappa^2}\left[\ln1.464+O\!\left(\frac{1}{(\kappa L)^4}\right)\right],
\label{eq:core-excess}
\end{equation}
does not depend on $L$: both the logarithms and the $1/[4(\kappa L)^2]$ terms cancel, and the
difference originates entirely in the core. This independence of the arbitrary radius $L$ is
the consistency check of the procedure. Substituting Eqs.~(\ref{eq:eps-v}) and
(\ref{eq:core-excess}) into Eq.~(\ref{eq:matching}) we find
\begin{equation}
\varepsilon=\frac{2\pi}{\kappa^2}\left[\ln(2\kappa)-\gamma+\ln1.464\right]=\frac{2\pi}{\kappa^2}\big(\ln\kappa+0.497\big),
\label{eq:eps-final}
\end{equation}
with corrections of order $\ln^{2}\kappa/\kappa^{4}$.  This reproduces the constant $0.497$ obtained from numerical solutions of the full GL
equations \cite{Hu} and identifies its two parts: $\ln2-\gamma$, supplied by the velocity
theory from the region outside the core, where Meissner screening replaces $\ln(\kappa L)$ of
the neutral problem by $\ln(2\kappa)-\gamma$; and $\ln1.464$, the excess energy of the GP core
over the normal core of the velocity theory.

\begin{figure}
\begin{center}
\includegraphics[width=1\columnwidth]{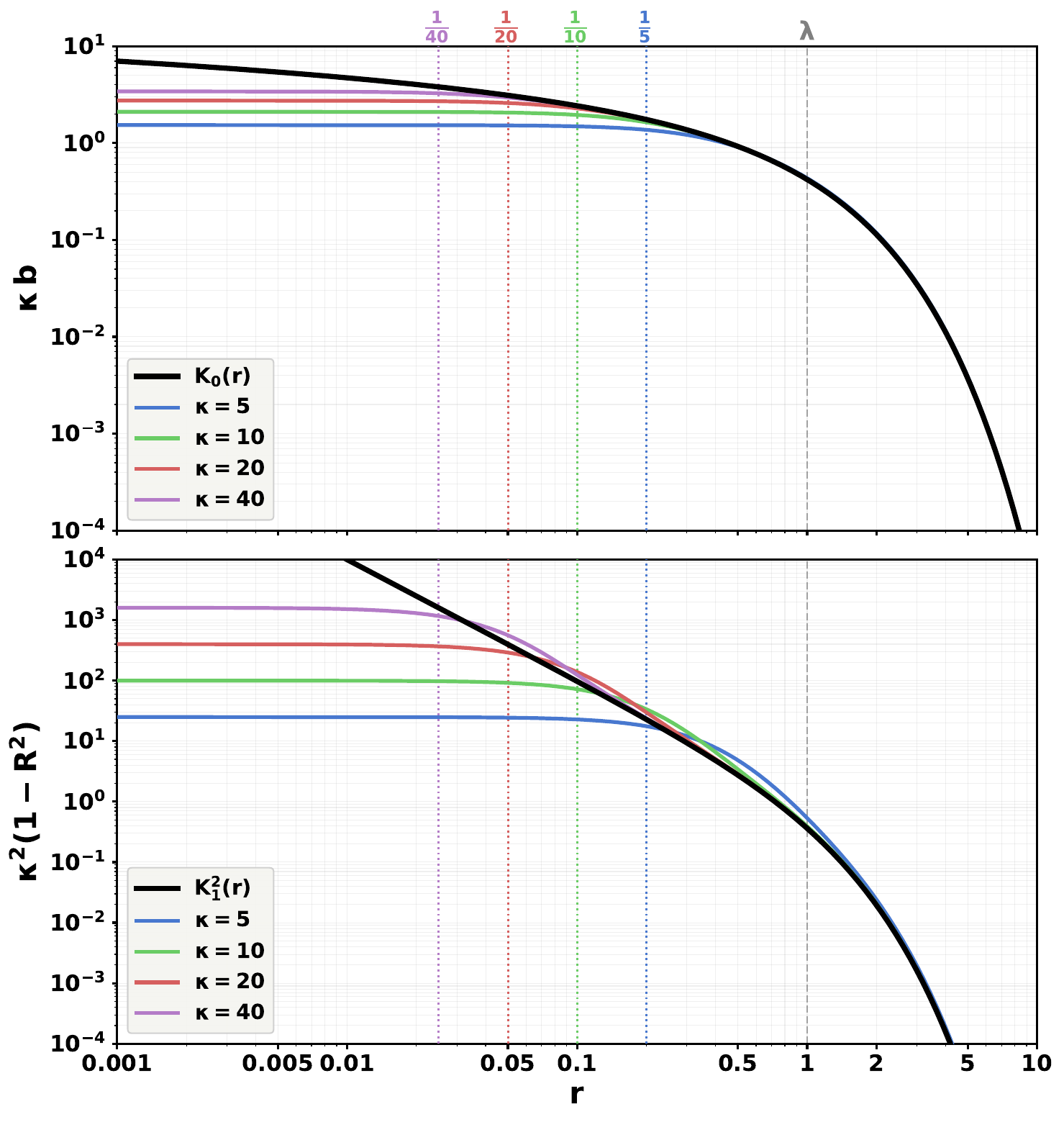}
\caption{Convergence of the outer $r>1/\kappa$ GL solutions of the Abrikosov vortex, Eqs.(\ref{GL1_vortex}), (\ref{GL2_vortex}), to the exact $\kappa\rightarrow\infty$ limit, Eqs.(\ref{magnetic_induction_exact}) and (\ref{superconducting_density_exact}), shown as solid black curves.}
\label{GL_Bessel_convergence}
\end{center}
\end{figure}

To illustrate the singular-limit structure, Eqs.(\ref{velocity_exact}), (\ref{magnetic_induction_exact}), and (\ref{superconducting_density_exact}), we solved Eqs.~(\ref{GL1_vortex}) and (\ref{GL2_vortex}) numerically subject to the boundary condition (\ref{singular_velocity}) for a sequence of increasing values of $\kappa$. Figure~\ref{GL_Bessel_convergence} compares the rescaled magnetic induction $\kappa b$ and the rescaled density deficit  $\kappa^{2}(1-R^{2})$ with the exact outer forms $K_{0}(r)$ and $K_{1}^{2}(r)$  implied by Eqs.~(\ref{magnetic_induction_exact}) and (\ref{superconducting_density_exact}). The core boundaries $r=1/\kappa$ are indicated by vertical dotted lines. As the core shrinks with increasing $\kappa$, the numerical GL solutions converge uniformly outside the core to the exact Bessel-function profiles.  

The figure also illustrates the algebraic slaving of the superconducting density to the velocity field in the $\kappa\rightarrow\infty$ limit. While the magnetic induction falls off as $K_{0}(r)\propto e^{-r}/\sqrt{r}$,
the density deficit behaves as $K_{1}^{2}(r)\propto e^{-2r}/r$ \cite{NIST} because it is quadratic in the velocity field.  This faster asymptotic decay, also observed in Figure \ref{GL_vortex_kappa20}, does not introduce an additional intrinsic length scale but follows directly from the exact constraint relating density and velocity (\ref{slaving_condition}).

Although the present work focuses on an isolated Abrikosov vortex, the statement of single-scale nature of the $\kappa\rightarrow\infty$ limit of GL theory is completely general. The conventional two-length-scale description of vortex interactions therefore applies only asymptotically close to the lower critical field, where vortex separations greatly exceed the penetration depth. Over a broad range of fields below the upper critical field, the extreme type-II regime requires a reformulation of the standard London-type description of the mixed state. The consequences of this reorganization will be explored elsewhere.
 
To conclude, we have shown that in the extreme type-II limit $\kappa\rightarrow\infty$, the Abrikosov vortex acquires a singular structure in which the GL equations reduce, outside a core shrinking as $\kappa\rightarrow\infty$, to a closed nonlinear theory for the superfluid velocity. The superconducting density becomes algebraically slaved to the velocity field, and the vortex exterior is determined exactly by a single length scale. The resulting Bessel-function type profile, Eqs.(\ref{velocity_exact}), (\ref{magnetic_induction_exact}), and (\ref{superconducting_density_exact}), is therefore not merely asymptotic but the exact outer solution of the limiting theory.

In this regime the conventional two-length-scale description no longer characterizes the outer vortex structure: although the intrinsic length scales remain formally distinct, the exterior configuration reorganizes into a single-scale object with all nonlinear spatial structure confined to a vanishing core. Unlike the London model, variations of the superconducting density are not frozen but retained through an exact algebraic constraint.

Because GL theory is equivalent to the static limit of the Abelian Higgs model \cite{Abelian}, the same strong-coupling limit yields an exactly solvable, single-scale Nielsen-Olesen string exterior. The extreme type-II limit thus reveals a hidden exact structure of one of the fundamental defects of superconductivity and gauge field theory.

The numerical computations and figures were produced with the assistance of Claude (Anthropic);  the code is available online \cite{GitHub}.


\begin{thebibliography}{18}

\bibitem{Abrikosov1}  A. A. Abrikosov, \textit{On the Magnetic Properties of Superconductors of the Second Group}, J, Exptl. Theoret. Phys. (U.S.S.R.) \textbf{32}, 1442 (1957) [Sov. Phys. JETP \textbf{5}, 1174 (1957)].

\bibitem{Abrikosov2}   A. A. Abrikosov, \textit{Fundamentals of the Theory of Metals} (Dover Publications, 2017), Part II.

\bibitem{de Gennes}  P. G. de Gennes, \textit{Superconductivity of Metals and Alloys},  (CRC Press, Boca Raton, 2018). 

\bibitem{LL9}  E. M. Lifshitz and L. P. Pitaevskii,  \textit{Statistical Physics}, Third edition, Part 2: Volume 9 (Course of Theoretical Physics) (Butterworth-Heinemann, 1980), Chapter V.

\bibitem{Abelian}  H. B. Nielsen and P. Olesen, \textit{Vortex-Line Models for Dual Strings}, Nucl. Phys. B \textbf{61}, 45 (1973).  

\bibitem{Rubakov}  V. A. Rubakov, \textit{Classical theory of gauge fields} (Princeton University Press, 2002).

\bibitem{Matricon}  P. G. de Gennes and J. Matricon, \textit{Collective Modes of Vortex Lines in Superconductors of the Second Kind}, Rev. Mod. Phys. \textbf{36}, 45 (1964).

\bibitem{Fetter}  A. L. Fetter and P. C. Hohenberg, \textit{Theory of Type II Superconductors}, in Superconductivity, vol 2, edited by R. D. Parks (Marcel Dekker, Inc., New York, 1969), pp.817-923.

\bibitem{Sarma}  D. Saint-James, G. Sarma, and E. J. Thomas, \textit{Type II Superconductivity}, (Pergamon Press, 1968).

\bibitem{BO}  M. Born and R. Oppenheimer, \textit{Zur Quantentheorie der Molekeln}, Annalen der Physik \textbf{389} (20), 457 (1927).

\bibitem{A}  P. W. Anderson, \textit{Plasmons, Gauge Invariance, and Mass},  Phys. Rev. \textbf{130}, 439 (1963).

\bibitem{H1}  F. Englert and R. Brout, \textit{Broken Symmetry and the Mass of Gauge Vector Mesons}, Phys. Rev. Lett. \textbf{13}, 321 (1964).

\bibitem{H2}  P. W. Higgs, \textit{Broken symmetries, massless particles and gauge fields}, Phys. Lett. \textbf{12}, 132 (1964).

\bibitem{H3} P. W. Higgs, \textit{Broken Symmetries and the Masses of Gauge Bosons}, Phys. Rev. Lett. \textbf{13}, 508 (1964).

\bibitem{H4} P. W. Higgs, \textit{Spontaneous Symmetry Breakdown without Massless Bosons}, Phys. Rev. \textbf{145}, 1156 (1966).  

\bibitem{H5} G. S. Guralnik, C. R. Hagen, and T. W. B. Kibble, \textit{Broken Symmetry and the Mass of Gauge Vector Mesons}, Phys. Rev. Lett. \textbf{13}, 585 (1964).

\bibitem{NIST} [DLMF] \textit{NIST Digital Library of Mathematical Functions}. \url{https://dlmf.nist.gov/}, Release 1.2.6 of 2026-03-15. F. W. J. Olver, A. B. Olde Daalhuis, D. W. Lozier, B. I. Schneider, R. F. Boisvert, C. W. Clark, B. R. Miller, B. V. Saunders, H. S. Cohl, and M. A. McClain, eds. Chapter 10, \textit{Bessel Functions}.

\bibitem{GP}  V. L.Ginzburg and L. P. Pitaevskii, \textit{On the Theory of Superfluidity}, ZhETF, \textbf{34}, 1240 (1958) [Sov. Phys. JETP, \textbf{7}, 858 (1958)].

\bibitem{Pethick}  C. Pethick and H. Smith, \textit{Bose-Einstein Condensation in Dilute Gases} (Cambridge University Press, 2002).

\bibitem{Hu}  C-R. Hu, \textit{Numerical Constants for Isolated Vortices in Superconductors}, Phys. Rev. B \textbf{6}, 1756 (1972).

\bibitem{GitHub}  \url{https://github.com/gkolomeisky/gl-vortex-semiclassical}.

\end{thebibliography}
\end{document}